# A Mechanization of the Blakers–Massey Connectivity Theorem in Homotopy Type Theory

Kuen-Bang Hou (Favonia) [*]

Carnegie Mellon University
favonia@cs.cmu.edu

Eric Finster [*]

LIX/École Polytechnique
ericfinster@gmail.com

Daniel R. Licata [*]

Wesleyan University
dlicata@wesleyan.edu

Peter LeFanu Lumsdaine [*]

Stockholm University
p.l.lumsdaine@gmail.com

## Abstract

This paper continues investigations in "synthetic homotopy theory": the use of homotopy type theory to give machine-checked proofs of constructions from homotopy theory.

We present a mechanized proof of the Blakers–Massey connectivity theorem, a result relating the higher-dimensional homotopy groups of a pushout type (roughly, a space constructed by gluing two spaces along a shared subspace) to those of the components of the pushout. This theorem gives important information about the pushout type, and has a number of useful corollaries, including the Freudenthal suspension theorem, which has been studied in previous formalizations.

The new proof is more elementary than existing ones in abstract homotopy-theoretic settings, and the mechanization is concise and high-level, thanks to novel combinations of ideas from homotopy theory and type theory.

## 1. Introduction

### 1.1 Homotopy-theoretical Aspects of Types

Types in Martin-Löf type theory can express an infinite-dimensional structure that corresponds to the notion of *space* studied in homotopy theory or the notion of an ∞-*groupoid* studied in higher category theory (Hofmann and Streicher 1998; Voevodsky 2006; Lumsdaine 2009; van den Berg and Garner 2011; Awodey and Warren 2009; Warren 2008; Gambino and Garner 2008). From this point of view, types in type theory have not only elements, but also paths (represented by the identity type $\mathsf{Id}_A(a,b)$), paths-between-paths (represented by the iterated identity type $\mathsf{Id}_{\mathsf{Id}_A(a,b)}(p,q)$), and so on. These higher paths have a complex algebraic theory, and characterizing them is a rich mathematical question known as calculating the *(higher) homotopy groups* of a space. For example, no feasible general method is known for calculating the higher homotopy groups of higher-dimensional spheres, which are in a sense the simplest ∞-groupoids, the $n$-sphere being freely generated by one (non-trivial) loop in dimension $n$. The source of this complexity is that paths in lower levels can give rise to non-trivial paths in higher levels.

At the first level, 1-dimensional paths have identity, composition, and inverse operations given by reflexivity, symmetry, and transitivity of equality. 2-dimensional paths (paths between paths) have identity, composition, and inverses, but also have additional algebraic structure induced by the 1-dimensional structure. For example, in addition to the usual composition · given by transitivity of equality, there is an additional "horizontal" composition that takes 2-paths $r : \mathsf{Id}_{\mathsf{Id}_A(a,b)}(p,q)$ and $s : \mathsf{Id}_{\mathsf{Id}_A(b,c)}(p',q')$ and produces a 2-path $\mathsf{Id}_{\mathsf{Id}_A(a,c)}(p \cdot p', q \cdot q')$, given by compatibility of composition. This additional composition operation in fact makes higher homotopy groups always commutative, even though 1-paths may not be. Next, the 2-dimensional paths induce further structure on the 3-dimensional paths, and so on. Though all this algebraic structure arises from the $J$ elimination rule for the identity type, it often requires substantial mathematical work to find an explicit description of the homotopy groups of a specific type.

A recent line of work on *synthetic homotopy theory* has investigated the path structures of particular higher inductive types (Lumsdaine and Shulman 2013; Shulman 2011a; Lumsdaine 2011) in type theory extended with Voevodsky's univalence axiom (Voevodsky 2006; Kapulkin et al. 2012). Using these tools, some elementary constructions from algebraic topology or homotopy theory have been developed and formalized using the proof assistants Agda (Norell 2007) and Coq (Coq Development Team 2009). These include calculations of some homotopy groups of spheres (Licata and Shulman 2013; Licata and Brunerie 2013; The Univalent Foundations Program 2013) and some homotopy equivalences (Licata and Brunerie 2015); constructions of the Hopf fibration (The Univalent Foundations Program 2013, §8.5), of covering spaces (Hou (Favonia) 2014), and of Eilenberg–Mac Lane spaces (Licata and Finster 2014); and proofs of the Freudenthal suspension theorem (The Univalent Foundations Program 2013, §8.6), the van Kampen theorem (The Univalent Foundations Program 2013, §8.7), and the Mayer–Vietoris theorem (Cavallo 2014). These developments are interesting from a formalization perspective because using the homotopy structure of types results in short, clean mechanized proofs. Moreover, they are interesting from a mathematical perspective because they are new proofs that com-bine ideas and techniques from homotopy theory and from type theory. Additionally, because the synthetic proofs abstract from the concrete details of a setting like topological spaces, they can be interpreted semantically in other higher-dimensional settings for homotopy theory that model homotopy type theory (Shulman 2013, 2015; Gambino and Sattler 2015)—so they prove many theorems at once by translation.

[*] This material is based on research sponsored in part by the National Science Foundation under grant numbers CCF-1116703, DMS-1128155, W911-NF0910273 and W911-NF1310154, by the Institute for Advanced Study's Oswald Veblen fund, and by the United States Air Force Research Laboratory under agreement numbers FA-95501210370 and FA-95501510053. The U.S. Government is authorized to reproduce and distribute reprints for Governmental purposes notwithstanding any copyright notation thereon. The views and conclusions contained herein are those of the authors and should not be interpreted as necessarily representing the official policies or endorsements, either expressed or implied, of the United States Air Force Research Laboratory, the U.S. Government or Carnegie Mellon University.

## 1.2 The Blakers–Massey Connectivity Theorem

In this paper, we present a formalization using the Agda proof assistant of a result called the Blakers–Massey connectivity theorem (Blakers and Massey 1951, 1952, 1953) (see (May 1999, Ch. 11, Section 1) for an introduction). This theorem concerns the *pushout* type $A +_C B$, which is specified by two types $A$ and $B$ and a dependent type $C : A \to B \to \mathsf{Type}$. The pushout is a generalization of a disjunction $A \vee B$ or disjoint union $A + B$, where certain "left" and "right" elements are identified. As a higher inductive type, it is specified by the following generators:

$$\begin{aligned}
\mathsf{left} &: A \to A +_C B \\
\mathsf{right} &: B \to A +_C B \\
\mathsf{glue} &: (a:A)(b:B) \to C\,a\,b \to \mathsf{Id}(\mathsf{left}(a), \mathsf{right}(b))
\end{aligned}$$

Like a disjoint union, we have left and right constructors. The new ingredient is that there is a path from $\mathsf{left}(a)$ to $\mathsf{right}(b)$ whenever $C$ relates $a$ and $b$. Because the pushout allows "gluing together" smaller spaces to make a bigger one, it plays a fundamental role in constructing spaces. Many constructions used in previous formalizations, such as the *suspension* type that was used to build higher spheres (The Univalent Foundations Program 2013, §6.5) and Eilenberg–Mac Lane spaces (Licata and Finster 2014), are special cases of pushouts (see The Univalent Foundations Program (2013, §6.8) for more examples). Quotients of disjoint unions also have applications in programming—for example, the integers can be described as the pushout of two copies of the natural numbers whose 0's are glued together. The first homotopy group of a pushout is characterized by the Seifert–van Kampen theorem, which was developed in homotopy type theory by Michael Shulman and formalized by Hou (Favonia) (see The Univalent Foundations Program (2013, §8.7)). This theorem says (roughly) that every path in the pushout can be decomposed as an alternating sequence of paths in $A$ and paths in $B$ whose middle points are related by $C$.

However, because of the complex algebraic structure of higher path spaces, it is more difficult to calculate the higher homotopy groups of a pushout—there is no straightforward general characterization. The Blakers–Massey theorem gives some partial information, which characterizes the higher homotopy groups in a certain range, depending on a property of $A$, $B$, and $C$, called *connectivity*. We will give a precise definition of connectivity below, but intuitively, a type is $n$-connected if its $k$-paths for $k \leqslant n$ are trivial. For example, the (2-dimensional) sphere is 1-connected, because any 1-dimensional loop can be filled in; but it is not 2-connected, because there are non-trivial 2-dimensional loops, which correspond to "going around the surface of the sphere." The notion of connectivity is extended to functions in such a way that, when a function $f : A \to B$ is $n$-connected, it induces an isomorphism on homotopy groups at level $k \leqslant n$ (and a surjection on the $(n+1)^{st}$ homotopy group)—so, intuitively, an $n$-connected function witnesses that $B$ is "trivial relative to $A$" in dimensions up to $n$, in the sense that up to this level it has no more information than $A$.

For any $a : A$, we can form the type $\Sigma b{:}B.\,C\,a\,b$ of "$B$'s that are related to $a$ by $C$," and similarly for any $b : B$ we can form the type $\Sigma a{:}A.\,C\,a\,b$ of "$A$'s that are related to $b$ by $C$." The Blakers–Massey theorem says that if for every $a : A$, $\Sigma b{:}B.\,C\,a\,b$ is $n$-connected, and if for every $b : B$, $\Sigma a{:}A.\,C\,a\,b$ is $m$-connected, then for every $a$ and $b$ the pushout constructor $\mathsf{glue}_{a,b} : C\,a\,b \to \mathsf{Id}(\mathsf{left}(a), \mathsf{right}(b))$ is $(n+m)$-connected. In particular, this implies that the homotopy groups of the path type $\mathsf{Id}(\mathsf{left}(a), \mathsf{right}(b))$ in the pushout (which are precisely the higher homotopy groups of the pushout itself) are the same as the homotopy groups in $C\,a\,b$ up to a certain point. Since information about the homotopy groups of the gluing family $C$ chosen to make a pushout is often known, this is a useful way to obtain information about the homotopy groups of the pushout itself. For example, it has as a special case the Freudenthal suspension theorem, which was used in past formalizations to calculate the $n^{th}$ homotopy group of the $n$-dimensional sphere (The Univalent Foundations Program 2013, §8.6) and to verify the correctness of a construction of Eilenberg–Mac Lane spaces (Licata and Finster 2014), and implies stability of the homotopy groups of spheres (in a certain range, increasing both the dimension of the sphere and the homotopy group by one gives the same group).

Relative to the existing work on synthetic homotopy theory, the computer-checked proof of the Blakers–Massey theorem that we report on in this paper is notable for a couple of reasons. First, while not a new result in homotopy theory, it seems to be an essentially new proof. Precisely, the present authors have translated this type-theoretic proof into a proof of Blakers–Massey in arbitrary ∞-toposes[1] (Finster et al. 2013, in preparation), significantly more elementary than previous proofs of the theorem in such settings (e.g. (Rezk 2010, Prop. 8.16)). Independently, based on the Agda formalization presented here, the homotopy theorist Charles Rezk has given another such translation (Rezk 2015); and Anel et al. (2016) have obtained new homotopy-theoretic results using generalizations of these translated proofs.

A second notable aspect of this proof is that it combines techniques from homotopy theory and from type theory in an interesting way. In developing synthetic homotopy theory, we have found that the same result can often be proved in ways that feel more "type-theoretic" or in ways that feel more "homotopy-theoretic"—for example, when a standard mathematical argument is translated to type theory and viewed from a perspective that emphasizes computation and normal forms and cut-free proofs, it sometimes looks indirect or reducible. The present work is the result of a back-and-forth between category theorists and type theorists, beginning with a calculation of the fundamental group of the circle by Shulman (Shulman 2011b), which was "reduced" by Licata (Licata and Shulman 2013), which led to a "type-theoretic" calculation of the diagonal homotopy groups of spheres by Licata and Brunerie (Licata and Brunerie 2013), which led to a calculation by Lumsdaine of the same using a more classical approach (the Freudenthal suspension theorem) (The Univalent Foundations Program 2013, §8.6), which led to the present result. To illustrate this interplay of type theory and homotopy theory, we will actually describe two different mechanizations of the Blakers–Massey theorem, one that is more direct in type theory, but involves some calculations that are hard to phrase in traditional homotopy-theoretic terms, and one that would be more familiar to a homotopy theorist, but is less direct, in the sense that it makes use of some intermediate types that are eliminated in the first version. Each mechanization is about 700 lines of code, on top of library code that is not specific to this theorem.

The remainder of this paper is organized as follows. In Section 2, we present some background definitions. In Section 3, we present the first mechanization of the Blakers–Massey theorem. In

---

[1] An ∞-*topos* is an abstract higher-categorical setting for homotopy theory, including structure corresponding to the structure present on spaces or ∞-groupoids; the prototypical ∞-topos is spaces, in the same way that the prototypical 1-topos is the category of sets. It is expected that some variant of homotopy type theory should have models in all ∞-toposes, but making this precise is an open question.

Section 4, we present the second version of the proof. Both proofs have been fully formalized in Agda.[2]

## 2. Background

We assume the reader is familiar with the basics of homotopy type theory, including higher inductive types and some synthetic calculations of homotopy groups (for introductions, see Licata and Shulman (2013); Licata and Finster (2014); The Univalent Foundations Program (2013)). In this section, we will review a few of the notions that play an important role in this paper. For the text of the paper, we will mostly use Agda syntax, but we will take some liberties that improve the presentation, which we comment on as we use them.

*Paths* We write $x = y$ as the identity type, the type of paths from x to y, refl as the constant path, ! p as the inverse of p, and p • q as the concatenation of the paths p and q.

*n-types* One of the homotopy-theoretic ideas adopted in homotopy type theory is that types are naturally stratified by the level (if any) at which their points, paths, paths-between-paths, etc. become trivial. For example, a proposition is a type A for which you can prove $(x\,y : A) \to x = y$ (points are trivial); a set, or 0-type is a type A with a proof of $(x\,y : A)\,(p\,q : x = y) \to p = q$ (paths are trivial). In general, we say that a type A is a $(-2)$-type (or is contractible) if there is an $x : A$ such that $(y : A) \to x = y$, and that A is an $(n+1)$-type if for all $x\,y : A$, $(x = y)$ is an n-type. See (The Univalent Foundations Program 2013, Ch. 3) for more details. In Agda, we write is-truncated n A to mean that A is an n-type, and n -Type for the (larger) type of all n-truncated types. We suppress the projection from n -Type to the universe, i.e. we treat an n -Type as a type.

*Truncations* Given any type A, there is a type τ n A, the *n-truncation* of A, which is "the best approximation of A by an n-type". This type has a constructor proj : A → τ n A, and an axiom stating that it is n-type. The recursion principle says that to define a function f : τ n A → C, where C is an n-type, it suffices to give a function b : A → C—and then f (proj x) ≡ b (x). More generally, the induction principle says that to define a function f : (x : τ n A) → C (x) where C : τ A n → n -Type is a family of n-types, it suffices to give b : (x : A) → C (proj x)—with the same computation rule. Truncations can be implemented using higher inductive types (The Univalent Foundations Program 2013, §6.9), so they need not be an extra primitive ingredient, but it is conceptually helpful to think of them as a type constructor. In (The Univalent Foundations Program 2013, §6.9), τ n A is written $||A||_n$ and proj is written $|-|$.

*Connected types* A dual notion to truncatedness is connectivity: an n-connected type is trivial at and *below* dimension n. Precisely, this may be defined using truncation (The Univalent Foundations Program 2013, Def. 7.5.1): A type A is *n-connected* iff τ n A is contractible. The n-truncation "kills" all the paths at levels above n, so if this is contractible, then the original type was trivial in dimensions ⩽ n. For example, a type A is 0-connected when the set-truncation of A is contractible—i.e. A has just one path-component. A type is 1-connected when its 1-truncation is contractible—i.e. it has one path-component, and trivial first homotopy group. In Agda, we write is-connected n A to mean that A is n-connected.

This definition of n-connected types is equivalent to a universal property, which says that "the type looks like a single point to every family of n-types." In type theory, this is expressed by the following induction principle:

connected-ind : ∀ { n A } { a0 : A } (P : A → n -Type)
  → is-connected (n + 1) A
  → (P a0)
  → ((x : A) → (P x))
connected-ind { a0 } P c p a0 = p

This says that if P is a family of n-types, then an element of P a0 for some a0 extends to an element of P x for all x, with a homotopy showing that on a0 this function gives back the original p.

*Connected maps* To a first approximation, n-connectivity of a *function* f : A → B is supposed to mean that the homotopy structure of B is the same as A up to level n, and thus that B is "trivial relative to A" up to n. As a first cut, one expects such a function f : A → B to induce an isomorphism on the homotopy groups of A and B up to level n, or on the n-truncations of A and B. The actual definition is slightly stronger, and implies both of these.

First, recall that f is an *equivalence* just if for each b:B, the *homotopy fiber* type Σ (a:A) f a = b (which we sometimes write as hfiber f b) is contractible.[3] The homotopy fiber is a proof-relevant preimage type: for each b, it consists of a point in a, together with a chosen path from where f sends that point to b. So saying each fiber is contractible is just a proof-relevant interpretation of "for each b:B, there is a unique a:A that f sends to b," a familiar definition of bijection.

The definition of n-connectivity that we use[4] is similar to this definition of equivalence, except it demands only that the n-truncation of the homotopy fiber is contractible:

is-connected-map n f = (b : B) → is-contr (τ n (Σ (a:A) f a = b))

Just as connectivity of types corresponds to an induction principle, so does connectivity of functions (The Univalent Foundations Program 2013, Lemma 7.5.7). The above definition of connectivity is equivalent to saying that every family of n-types sees B as being generated by A, in the sense that to prove P b for all b, it suffices to prove P (f a) for all a:

connected-map-ind : ∀ { n A B } (f : A → B) (P : B → n -Type)
  → is-connected-map n f
  → ((a : A) → P (f a))
  → ((b : B) → P b)
connected-map-ind f P c f p (f a) = p a

There are a couple of ways to gain intuition for this definition of connectivity. First, setting aside the truncation, the definition looks like the concept of surjection—for all b, there exists an a that maps to b. Indeed, a function is $(-1)$-connected iff it is surjective, which is best rendered in type theory as saying that there merely exists a preimage of every b using the $(-1)$-truncation (The Univalent Foundations Program 2013, Lemma 7.5.2). In general, this definition corresponds to surjectivity on homotopy groups up to dimension n+1—for example, if f is 0-connected, then one can show that it is surjective on $\pi_1$, the group made from paths, and $\pi_0$, the set of connected components. Moreover, if a function is surjective on homotopy group k+1, it is injective on the homotopy group k. For example, when f is 1-connected, it is a surjection on $\pi_1$, which implies that for every path f a0 = f a1, there merely exists a path a0 = a1, which is very close to the definition of injectivity on $\pi_0$. A function that is both surjective and injective (and a homomorphism) is an isomorphism, so when f is n-connected it induces an isomorphism on homotopy groups up to n. Therefore, it implies our first cut at the notion of connectivity, and additionally specifies surjection on the n+1th homotopy group. These conditions together—isomorphism up to $\pi_n$, and surjectivity on $\pi_{n+1}$—comprise the standard homotopy-theoretic definition.

---

[3] We write Σ (x:A) B for dependent pair types, though it is not valid Agda.

[4] In our main Agda library we call this has-connected-fibers, to distinguish it among other equivalent definitions of connectivity.

Another intuition is that an n-connected map f : A → B is an extension of A obtained just by adjoining cells of dimension greater than n+1. For instance, a (−1)-connected extension A → B may add paths between existing points of A (and higher paths similarly), but may not add any new points, and hence must remain a surjection on 0-truncations. Under this viewpoint, the induction principle for connected maps follows, because when mapping into an n-truncated goal, once the action on A is known, images for the adjoined (n+1)-cells of B are automatically (and uniquely) given.

One word of warning: there exist several different conventions for the indexing of connectivity. Many texts use (n+1)-connected where we use n-connected; the relationship between the indexing for spaces and maps also varies by ±1.

*Pushouts*  For two types X and Y, and Q : A → B → Type, the pushout type Pushout X Y Q is a higher inductive type with constructors

    left   : X → Pushout X Y Q
    right  : Y → Pushout X Y Q
    glue   : {x : X} {y : Y} → Q x y → left x = right y

The (non-dependent) recursion principle is

    pushout-rec-nondep : ∀ R
      (left* : X → R)
      (right* : Y → R)
      (glue* : ∀ {x} {y} (q : Q x y) → left* x = right* y)
      → (Pushout X Y Q → R)
    pushout-rec-nondep R left* right* glue* (left x) ≡ left* x
    pushout-rec-nondep R left* right* glue* (right x) ≡ right* x
    ap (pushout-rec-nondep R left* right* glue*) (glue q) = glue* q

This is like case-analysis for a disjoint union X + Y, except that you also need to show that the image of left and the image of right agree on Q. The computation rules on points are definitional equalities, while on paths it is a propositional equality.

The full dependent induction principle is

    pushout-rec : ∀ (R : Pushout X Y Q → Type)
      (left* : (x : X) → R (left x))
      (right* : (y : Y) → R (right y))
      (glue* : ∀ {x} {y} (q : Q x y)
              → (transport R (glue q) (left* x)) = right* y)
      → ((p : Pushout X Y Q) → R p)

and it has analogous computation rules.

*Wedge Connectivity*  One particular use of the pushout is the *wedge* of two pointed spaces. Suppose X and Y are types with x0 : X and y0 : Y. Then the wedge X ∨ Y is the pushout of X and Y with x0 glued to y0:

    X ∨ Y = Pushout X Y (λ x y → x = x0 × y = y0)

Informally, if X and Y are both the real line, and x0 and y0 are both 0, then X ∨ Y can be thought of as the axes of the plane. Then the function that "includes the axes into the plane" may be defined as follows:[5]

    wedge-to-product : X ∨ Y → X × Y
    wedge-to-product (left x) = (x , y0)
    wedge-to-product (right y) = (x0 , y)
    ap wedge-to-product (glue (refl , refl)) = refl

Thinking of left x as "a point on the *x*-axis", it is sent to the pair (x, y0), and similarly for right y. In the final clause, after some path induction, we must show that wedge-to-product (left x0) = wedge-to-product (right x0), which is true by definition.

---
[5] Formally, this is an application of pushout-rec-nondep, but for readability we use a clausal notation.

One of the key lemmas used in the proof of the Blakers–Massey theorem concerns the connectivity of this map wedge-to-product: If X is m-connected and Y is n-connected, then wedge-to-product is (m+n)-connected. In the proof in Section 4, we will use this fact, and it is proved in the formalization. For intuition, take X and Y to be $\mathbb{S}^2$, the usual 2-dimensional sphere, which is 1-connected, because any path can be contracted. Then this lemma says that up to and including their 2-dimensional paths, $\mathbb{S}^2 \times \mathbb{S}^2$ and $\mathbb{S}^2 \vee \mathbb{S}^2$ have the same structure. The second homotopy group of $\mathbb{S}^2 \times \mathbb{S}^2$ is $\mathbb{Z} \times \mathbb{Z}$ because $\pi_2(\mathbb{S}^2) = \mathbb{Z}$ and homotopy groups commute with products. $\mathbb{S}^2 \vee \mathbb{S}^2$ can be visualized as two spheres joined at a point. A 2-dimensional path is a homotopy from the constant path at this point to itself, which can either go around the left sphere, around the right sphere, or around both at the same time. But these two can be commuted, because we can always pause the movement at one side while finishing that on the other side. So it makes sense that its second homotopy group is also $\mathbb{Z} \times \mathbb{Z}$.

In the proof in Section 3, we will use this fact in a bit of a disguised way: There are no wedges in the statement of the Blakers–Massey theorem, so from a proof theoretic perspective, it seems a bit indirect to introduce a wedge type just to use this lemma. We can avoid this detour by phrasing the connectivity of the inclusion of the wedge into the product as a derived induction principle, by combining it with connected-map-ind and with pushout-rec for the wedge. When we put these ingredients together, we get the following principle:

    wedge-connect : ∀ {m n p} {A B : Type} {a_0 : A} {b_0 : B}
      (cA : Connected m A) (cB : Connected n B)
      (C : A → B → m+n -Types)
      (fa_0 : (b : B) → C a_0 b)
      (fb_0 : (a : A) → C a b_0)
      (fab_0 : fa_0 b_0 = fb_0 a_0)
      (a : A) (b : B) → C a b
    α : wedge-connect [...] fa_0 fb_0 fab_0 a_0 b = fa_0 b
    β : wedge-connect [...] fa_0 fb_0 fab_0 a b_0 = fb_0 a
    [... α b_0 and β a_0 agree up to fab_0 ...]

This says that if A is m-connected and B is n-connected, with points $a_0$ and $b_0$, and C is a family of (m+n)-types, then to construct C for all a and b, it suffices to construct it for all b, fixing $a_0$, and for all a, fixing $b_0$, in such a way that these two constructions agree when both are constructing C $a_0$ $b_0$. Additionally, there are "computation" homotopies saying that on $a_0$ and $b_0$, the overall construction behaves like what we put in (and a third homotopy saying that these two are coherent). This principle was developed by Lumsdaine to prove the Freudenthal suspension theorem (a special case of Blakers–Massey) in type theory; for a proof, see (The Univalent Foundations Program 2013, Lemma 8.6.2).

## 3. First Proof

### 3.1 Formulation

With the definitions of the previous section, we can now state the Blakers–Massey theorem.

**Theorem.** *Let X and Y be types, and Q a family X → Y → Type. Suppose m,n ⩾ −1, and for each x:X the type Σ (y:Y) (Q x y) is m-connected, and dually for each y:Y the type Σ (y:Y) (Q x y) is n-connected.*

*Then for each x:X and y:Y, the map glue {x} {y} : Q x y → left x = right y is (m+n)-connected, where left x = right y is a path type of Pushout X Y Q.*

For the rest of this section, fix X, Y, Q, m, n as in the theorem.

The connectivity hypotheses are usually stated as connectivities of the projection maps from the total space $\Sigma\,(x{:}X)\,(y{:}Y)\,(Q\,x\,y)$. The types in the present form can be seen as the fibers of these maps; or, more directly, $\Sigma\,(y{:}Y)\,(Q\,x\,y)$ can be read as "the type of y's that are related to x by Q", and $\Sigma\,(y{:}Y)\,(Q\,x\,y)$ dually.

The goal, which we now need to prove, relates the homotopy groups of the original family Q to the homotopy groups of the resulting pushout itself, up to dimension $m+n$. Unwinding the definition of connectivity of a map, it is the statement:

blakers-massey : $\{x_0 : X\}\,\{y : Y\}\,(r : \text{left}\,x_0 = \text{right}\,y)$
  $\to$ is-connected $(n+m)$ (hfiber glue r)

We single out $x_0$ with a subscript, since we will fix it throughout the rest of the proof. (Similarly, in the formalization, we take it as a section parameter.) Recalling that is-connected n X is defined as is-contr $(\tau\,n\,X)$, this unwound form of the theorem can be thought of as saying that for every path r : left $x_0$ = right y, there is an element q : Q $x_0$ y (in the domain of glue) that is a kind of explicit representation or canonical form for r, up to level $n{+}m$.

Overall, the (perhaps rather mysterious) connectivity hypotheses are used twice: once rather weakly, to supply some extra auxiliary assumptions, and once more substantially to apply the wedge connectivity lemma. It is the wedge connectivity lemma which gives rise to the additivity of connectivity in the conclusion.

### 3.2 Definition of code

To prove blakers-massey, we want to apply *based path induction* (The Univalent Foundations Program 2013, Lemma 7.5.2) on the path r, so that we only have to consider the case where r is reflexivity. However, this requires either left $x_0$ or right y to be generalized over the whole pushout. (The intuition is that any path can be contracted to the degenerate reflexivity path, provided at least one endpoint is able to vary freely.) So we want to generalize the original goal to the statement

blakers-massey' : $\{p : \text{Pushout}\,X\,Y\,Q\}\,(r : \text{left}\,x_0 = p)$
  $\to$ is-contr (code $\{p\}$ r)

for some family of types

code : $\{p : \text{Pushout}\,X\,Y\,Q\} \to (\text{left}\,x_0 = p) \to \text{Type}$

such that

code $\{\text{right}\,y_1\}$ r = $\tau\,(n+m)$ (hfiber glue r)

Recall that is-connected n X is defined as is-contr $(\tau\,n\,X)$, and so, with this definition of code for right, is-contr (code $\{\text{right}\,y_1\}$ r) is exactly the original goal. Just as the original goal gives a kind of canonical form for the n+m-level information in r when r is a path from inl to inr, so code $\{p\}$ r can be thought of as a type of explicit characterizations or canonical forms for (the n+m-level information in) a path r with an endpoint anywhere in the pushout.

The family code will be defined by recursion on p, so we need to feed these three components to the recursion principle of the pushout:

- code $\{\text{left}\,x_1\}$ (r : left $x_0$ = left $x_1$) for any $x_1$ : X; and
- code $\{\text{right}\,y_1\}$ (r : left $x_0$ = right $y_1$) for any $y_1$ : Y, defined as above to be $\tau\,(n+m)$ (hfiber glue r); and
- ap ($\lambda\,p \to$ code $\{p\}$) glue.

The difficulty is to find the analogue of the theorem for the left $x_1$ case, that is, when both the end points of r are in X. Our trick is to make our assumptions more symmetric, by supposing we have some distinguished $y_0$ : Y and $q_{00}$ : Q $x_0$ $y_0$ while defining code; like $x_0$, $y_0$ and $q_{00}$ will be fixed through most of the rest of the argument (using Agda's section mechanism, so we do not write them as arguments explicitly). We will show that we can discharge these extra assumptions below.

The list of three needed components for defining code remain the same, as does the definition in the right case, but for the other cases we will now make use of newly added arguments $y_0$ and $q_{00}$. In terms of diagrams, the above definition of the right case code $\{\text{right}\,y_1\}$ (r : left $x_0$ = right $y_1$), which was chosen to make our generalization imply the original theorem, can be drawn as follows.[6] It is the type of all $q_{01}$'s such that

$$\begin{array}{c}\text{left } x_0 \\ r \downarrow \qquad = \qquad \text{glue } q_{01} \searrow \\ \text{right } y_1 \qquad\qquad \text{right } y_1 \end{array} \quad (1)$$

For the left case, we define

code $\{\text{left}\,x_1\}$ (r : left $x_0$ = left $x_1$) =
  $\tau\,(n+m)$ (hfiber ($\lambda\,q_{10} \to$ glue $q_{00}\bullet$ ! (glue $q_{10}$)) r)

This represents (the truncation of) the type of all $q_{10}$'s such that

$$\begin{array}{c}\text{left } x_0 \quad\quad \text{left } x_0 \xrightarrow{\text{glue } q_{00}} \text{right } y_0 \\ r \downarrow \qquad = \qquad \swarrow\;!\,(\text{glue } q_{10}) \\ \text{left } x_1 \quad\quad \text{left } x_1 \end{array} \quad (2)$$

The remaining missing piece is ap ($\lambda\,p \to$ code $\{p\}$) glue, that is, to show the above types are equivalent when there is $q_{11}$ : Q $x_1$ $y_1$ such that glue $q_1$ connects left $x_1$ to right $y_1$. This boils down to an equivalence between the type code $\{\text{left}\,x_1\}$ "transported" along glue $q_1$, and the type code $\{\text{right}\,y_1\}$. Pictorially, the type code $\{\text{left}\,x_1\}$ after transportation maps each r : left $x_0$ = right $y_1$ to a type of all $q_{10}$'s such that

$$\begin{array}{c}\text{left } x_0 \quad\quad \text{left } x_0 \xrightarrow{\text{glue } q_{00}} \text{right } y_0 \\ r \searrow \qquad = \qquad \swarrow\;!\,(\text{glue } q_{10}) \\ \text{right } y_1 \quad\quad \text{left } x_1 \xrightarrow[\text{glue } q_{11}]{} \text{right } y_1 \end{array} \quad (3)$$

The goal is to show, for any r, there is an equivalence between the truncation of the type of all $q_{01}$'s satisfying Equation 1 and that of all $q_{10}$'s satisfying Equation 3. It turns out that this equivalence is non-trivial and heavily relies on the connectivity conditions in the Blakers–Massey.

We can slightly simplify the needed equivalence by eliminating the middle r; ignoring the truncation for the moment, essentially we want to prove that for any $q_{01}$ there is a $q_{10}$ such that the equation

$$\begin{array}{c}\text{left } x_0 \quad\quad \text{left } x_0 \xrightarrow{\text{glue } q_{00}} \text{right } y_0 \\ \text{glue } q_{01} \searrow \qquad = \qquad \swarrow\;!\,(\text{glue } q_{10}) \\ \text{right } y_1 \quad\quad \text{left } x_1 \xrightarrow[\text{glue } q_{11}]{} \text{right } y_1 \end{array} \quad (4)$$

is true, and vice versa. Then we must show that these two functions are inverse to each other, which establish an equivalence between all $q_{01}$'s and $q_{10}$'s. In this section we will only demonstrate the direction from $q_{10}$ to $q_{01}$ as the other is symmetric. Ignoring the truncations, we wish to show

$\forall\,(q_{10} : Q\,x_1\,y_0)$
  $\to \Sigma\,(q_{01} : Q\,x_0\,y_1)\,(\text{glue } q_{01} = \text{glue } q_{00} \bullet\,!\,(\text{glue } q_{10}) \bullet \text{glue } q_{11})$

---

[6] Truncations are ignored in diagrams.

The idea is to *reorder* all the hypotheses (including abstracting over the fixed $x_0$, $y_0$ and $q_{00}$) to match the wedge connectivity theorem. After the reordering the lemma is

$\forall\, x_1\, y_0\, (q_{10} : Q\, x_1\, y_0)\, x_0\, (q_{00} : Q\, x_0\, y_0)\, y_1\, (q_{11} : Q\, x_1\, y_1)$
$\to \Sigma\, (q_{01} : Q\, x_0\, y_1)\, (\text{glue } q_{01} = \text{glue } q_{00} \bullet\, !\, (\text{glue } q_{10}) \bullet \text{glue } q_{11})$

In intuition, with the wedge connectivity theorem and proper truncations, we only have to consider the cases either $q_{00}$ collides with $q_{10}$ or $q_{11}$ collides with $q_{10}$, as long as we give a coherent choice when both collide.

Formally, while applying wedge-connect in Section 2, let A be the type of the upper arm, $\Sigma\, (x : X)\, (Q\, x\, y_0)$, B be the lower arm, $\Sigma\, (y : Y)\, (Q\, x_1\, y)$, and C be

$\lambda\, \{(x_1, q_{11})\, (y_0, q_{00}) \to$
$\tau\, (n + m)\, (\text{hfiber glue } (\text{glue } q_{00} \bullet\, !\, (\text{glue } q_{10}) \bullet \text{glue } q_{11}))\}$

The connectivities of A and B are exactly the original hypotheses of Blakers–Massey, and the truncation level of C is forced by the explicit truncation. $(x_1, q_{10})$ is a point of A and $(y_0, q_{10})$ is a point of B, which signifies the collision of either $q_{00}$ or $q_{11}$ with $q_{10}$. The term $fa_0$, which represents the case where the upper arm $q_{00}$ collides to the diagonal $q_{10}$, is the pair $(q_{11}, \text{lemma})$ where lemma is a path of type

$\text{glue } q_{11} = \text{glue } q_{10} \bullet\, !\, (\text{glue } q_{10}) \bullet \text{glue } q_{11}$

On the other hand, the term $fb_0$, which represents the case where the lower arm $q_{11}$ collides to the diagonal $q_{10}$, is the pair $(q_{11}, \text{lemma})$ where lemma is a path of type

$\text{glue } q_{00} = \text{glue } q_{00} \bullet\, !\, (\text{glue } q_{10}) \bullet \text{glue } q_{10}$

The last argument $fab_0$ is to show that $fa_0$ and $fb_0$ agree on $q_{10}$, which follows from the fact that $q_{10}$ is picked when both arms collide with the diagonal. In sum, these define a function from $q_{10}$'s to $q_{01}$'s.

The other direction can be defined similarly, and through the same technique one can show they are inverse to each other. This eventually fills the final piece of type

$\text{ap}\, (\lambda\, p \to \text{code}\, \{p\})\, \text{glue}$

and concludes the construction of code. In total, this construction is the largest part of the proof, and consists of approximately 500 lines of code. In the verification that the two maps are inverse, the mechanization involves some clever abstraction over portions of the proof that can be shared between both the $fa_0$ and $fb_0$ cases of wedge-connect lemma, which simplifies showing the final $fab_0$ coherence assumption of this lemma.

It now remains to show the contractibility of code $\{p\}$ r, for each p : Pushout X Y Q and r : left $x_0$ = p.

### 3.3 Contractibility of code

A straightforward way to show contractibility of a type A is to give a point center : A, and a *contraction* from A to this point, i.e. a path from every other point of A to center.

***Centers.*** So we want to give a point of code $\{p\}$ r, for each p : Pushout X Y Q and r : left $x_0$ = p. By based path induction on r, it is enough to give a point of code (left $x_0$) refl. By the construction of code, this type reduces to

$\tau\, (n + m)\, (\text{hfiber}\, (\lambda\, q_{10} \to \text{glue } q_{00} \bullet\, !\, (\text{glue } q_{10}))\, \text{refl})$

which is inhabited by the projected pair

$\text{proj}\, (q_{00}, \text{inv-r}\, (\text{glue } q_{00}))$

where inv-r path : path $\bullet$ ! path = refl is a proof that any path concatenated with its inverse is homotopic to refl.

Putting this together, we obtain the desired centers:

code-center : $\forall\, \{p\}\, r \to \text{code}\, \{p\}\, r$
code-center refl = proj $(q_{00}, \text{inv-r}\, (\text{glue } q_{00}))$

***Contractions.*** We now want a contraction on each code $\{p\}$ r, to the point code-center $\{p\}$ r. The type $\Sigma\, (p : \text{Pushout X Y Q})\, (\text{left } x_0 = p)$ of pairs of such p and r is contractible, so it is enough to give a contraction for any specific pair. We give it for the pair right $y_0$, glue $q_{00}$; but to do this, we step back to an intermediate generality, and show

$\forall\, \{y_1 : Y\}\, (r : \text{left } x_0 = \text{right } y_1)\, (c : \text{codes}\, \{y_1\}\, r)$
$\to c = \text{code-center}\, \{\text{right } y_1\}\, r$

i.e. the case where p is right $y_1$ for some $y_1$ : Y, and r is arbitrary. The re-generalization of r is needed for the path-induction below.

By construction, code $\{$ right $y_1$ $\}$ r is just $\tau\, (n + m)\, (\text{hfiber glue } r)$. Using the induction principles for truncations, hfiber (as pairs), and paths, we may assume that c is of the form proj $(q_{01}, \text{refl})$ for some $q_{01}$ : Q $x_0$ $y_1$, and r is glue $q_{01}$. So it remains to show

proj $(q_{01}, \text{refl})$ = code-center $\{$ right $y_1$ $\}$ (glue $q_{01}$)

This is a quite involved but mostly routine calculation, of about 50 lines; interested readers may check the Agda mechanization directly for details (code-coh-lemma).

This gives a contraction to code-center on code $\{$ right $y_1$ $\}$ r, for any $y_1$, r. In particular, it gives us a contraction on code $\{$ right $y_0$ $\}$ (glue $g_{00}$), and hence, by the contractibility of the type of pairs (p, r), on each type code $\{p\}$ r, as required for blakers-massey''.

### 3.4 Theorem

We have shown our second generalization assuming fixed $x_0$ : X, $y_0$ : Y, and $q_{00}$ : Q $x_0$ $y_0$ (which were implicit parameters in the above constructions). Making these explicit in the type of blakers-massey'', we have shown

blakers-massey'' : $\forall\, \{x_0\}\, \{y_0\}\, (q_{00} : Q\, x_0\, y_0)$
$\{p : \text{Pushout X Y Q}\}\, (r : \text{left } x_0 = p)$
$\to$ is-contr (code $x_0$ $y_0$ $q_{00}$ $\{p\}$ r)

We need to show that this implies the first generalization

blakers-massey' : $\forall\, \{x_0\}\, \{p : \text{Pushout X Y Q}\}\, (r : \text{left } x_0 = p)$
$\to$ is-contr (code $x_0$ $\{p\}$ r)

which, by definition of code, immediately implies the original blakers-massey by taking p of the form right y.

So the only remaining gap is the extra assumptions $y_0$ and $q_{00}$ in blakers-massey''. Given $x_0$ for blakers-massey', the remaining goal is a (−1)-type, and we assumed that m is at least −1, so the goal is *a fortiori* m-truncated. Our connectivity assumptions say that $\Sigma\, (y : Y)\, (Q\, x_0\, y)$ is m-connected, so in particular its m-truncation has an element. Because the goal is an m-type, we can eliminate the truncation to obtain some element of it; i.e. a pair $y_0$, $q_{00}$ as desired. This closes the gap and finishes the proof of the main theorem.

## 4. Second Proof

In this section, we describe a second mechanization of the Blakers–Massey theorem.[7] Qualitatively, this proof is "less type-theoretic" and "more homotopy-theoretic". For example, relative to Section 3, it introduces some intermediate steps (cut formulas) that are not used above. On the other hand, these intermediate steps allow more of the proof to be described in mathematical language, and there are fewer spots in the proof that rely on reductions/calculations that are difficult to convey outside of a proof assistant (though there are still some). In particular, the construction of the equivalences in the codes fibration has a higher-level description.

***Fat Pushouts*** For this section, it will be helpful to use an alternative, "fat", presentation of the pushout type. Suppose we have X

---

[7] This formalization uses a different Agda library, so we ask the reader to bear with some changes in naming and capitalization conventions.

and Y and P : X → Y → Type, then the fat pushout adds a "middle" constructor coming from P, and when p : P x y, rather than gluing inl x to inr y directly, it glues them both to inm p.

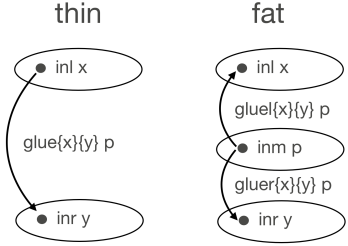

It is specified by the following constructors:

inl : X → Pushout X Y P
inr : Y → Pushout X Y P
inm : {x : X} {y : Y} → P x y → Pushout X Y P
gluel : {x : X} {y : Y} → (p : P x y) → (inm p) = (inl x)
gluer : {x : X} {y : Y} → (p : P x y) → (inm p) = (inr y)

and the usual elimination rules for such a higher inductive. This is equivalent to the previous presentation. For example, we define

glue : {x : X} {y : Y} (p : P x y) → (inl x) = (inr y)
glue p = gluer p ∘ ! (gluel p)

by composing gluel and gluer[8], going through the new middle point. The fatness does not add any information, roughly because we can contract away inm p and gluel p by path induction, leaving only gluer/glue. The fat pushout is helpful because it will allow us to break up the definition of the equivalences in the codes fibration in a more symmetric way.

***Properties of hfibers and connectivity***  We will use the following lemmas about homotopy fibers and connectivity.

First, when e is an equivalence, the homotopy fiber of f at e x (i.e. $\Sigma$ (a : A) f a = e x) is equivalent to the homotopy fiber of ($e^{-1}$ o f) at x (i.e. $\Sigma$ (a : A) $e^{-1}$ (f a) = x), by moving the equivalence to the other side of the equation:[9]

hfiber-at-equiv : ∀ { A B B' } (f : A → B') (e : B ≃ B') (x : B)
  → HFiber f (e x) ≃ HFiber ($e^{-1}$ o f) x

Second, suppose we have a "fiberwise" function f : (x : A) → B x → C x. This induces a function

(λ (a',b') → (a' , f a' b')) : ($\Sigma$ (a : A) B a) → ($\Sigma$ (a : A) C a)

on the *total spaces* of B and C, which are the $\Sigma$-types that package elements of B and C together with the A that they depend on. If we have a particular a : A and c : C a, then the fiber of the partial application f a at c is the same as the fiber of the total space at (a,c):

HFiber-fiberwise-to-total-eqv : {A : Type} {B C : A → Type}
  {a : A} {c : C a} (f : (x : A) → B x → C x)
  → (HFiber (f a) c) ≃ (HFiber (λ (a',b') → (a' , f a' b')) (a , c))

Roughly, the right-hand type consists of an (a',b') : ($\Sigma$ (a : A) B a) such that (a', f a' b') = (a,c), so contracting the equality between the first components of these pairs shows that it is the same as the left-hand type.

Another fact is that this induced map on the total space is n-connected when f is n-connected for each element of A:[10]

fiberwise-to-total-connected : ∀ { n A } {B C : A → Type}
  (f : (x : A) → B x → C x)

---
[8] $\beta \circ \alpha$ is path concatenation in function composition order; i.e. $\alpha : x = y$ $\beta : y = z$.
[9] We elide the projection from an equivalence to the underlying function.
[10] ConnectedMap is what was called is-connected-map in Section 3.

  → ((x : A) → ConnectedMap n (f x))
  → ConnectedMap n (λ (a',b') → (a' , f a' b'))

See (The Univalent Foundations Program 2013, Lemma 7.5.13) for a proof.

Connectivity is preserved by pre- and post- composition with an equivalence:

∀ { n A B B' } {f : A → B} (e : B ≃ B')
  → ConnectedMap n f → ConnectedMap n (e o f)
∀ { n A A' B } {f : A → B} (e : A' ≃ A)
  → ConnectedMap n f → ConnectedMap n (f o e)

An n-connected map f : A → A' induces an equivalence on the truncations Trunc n A ≃ Trunc n A'.[11] Below, we will need a condition under which this equivalence extends to the truncations of the fibers of maps g : A → B and h : A' → B, i.e. the truncations of $\Sigma$ (a:A) g a = b and $\Sigma$ (a':A') h a' = b. We prove in the formalization that it is sufficient to ask that h o f = g.

fiber-top-equiv : ∀ { n A A' B } (f : A → A') (g : A → B) (h : A' → B)
  → ConnectedMap n f
  → ((x : A) → h (f x) = g x)
  → { b : B } → (Trunc n (HFiber g b)) ≃ (Trunc n (HFiber h b))

### 4.1 Construction of CodesFor

For the remainder of this proof, we fix X Y : Type and P : X → Y → Type such that

(x : X) → Connected i ($\Sigma$ (y : B) P x y) where j ⩾ −1
(y : Y) → Connected j ($\Sigma$ (x : A) P x y) where j ⩾ −1

We write W for the fat Pushout X Y P, and Z for $\Sigma$ ((x , y) : X × Y) P x y.

For this subsection, we fix an element of Z, written ($x_0$:X, $y_0$:Y, $p_0$ : P $x_0$ $y_0$). Our first goal is to construct

CodesFor : (w : W) ($\alpha$ : (inm $p_0$) = w) → Type

Like above, CodesFor w $\alpha$ will be a type of explicit characterizations or canonical representations of the path $\alpha$. We define CodesFor by pushout-elimination on w, which requires 5 cases. First, for inl x, we need to give a type dependent on x:X and $\alpha$ : inm $p_0$ = inl x; for inr y, a type dependent on y:Y and $\alpha$ : inm $p_0$ = inr y; and for inm p, a type dependent on x:X, y:Y, p:P x y, and $\alpha$ : inm $p_0$ = inm p. Then for gluel and gluer we need to construct some equivalences between the types given in the first step.

***Fibers***  For the inl and inr cases, given any x related to $y_0$, or any y related to $x_0$, we can define can define paths from inm $p_0$:

$gluel_0$ : {x : X} → P x $y_0$ → (inm $p_0$) = (inl x)
$gluel_0$ $pxy_0$ = ! (glue $pxy_0$) ∘ gluer $p_0$
$gluer_0$ : {y : Y} → P $x_0$ y → (inm $p_0$) = (inr y)
$gluer_0$ $pxy_0$ = glue $pxy_0$ ∘ gluel $p_0$

For example, $gluel_0$ produces the composite

inm $p_0$ = inr $y_0$ = inm $pxy_0$ = inl x

while $gluer_0$ goes through inl $x_0$. We define the cases of the codes fibration for inl and inr as the truncations of the homotopy fibers of these maps:

CodesFor (inl x) ($\alpha$ : inm $p_0$ = inl x) = Trunc i+j (HFiber $gluel_0$ $\alpha$)
CodesFor (inr y) ($\alpha$ : inm $p_0$ = inr y) = Trunc i+j (HFiber $gluer_0$ $\alpha$)

For example, this says that the type of "normal forms" for $\alpha$ : inm $p_0$ = inl x is the (i+j)-truncation of $\Sigma$ ($pxy_0$ : P x $y_0$) $gluel_0$ p = $\alpha$. The $\Sigma$-type represents a link $pxy_0$ from x to $y_0$, such that the path in the pushout given as above is homotopic to $\alpha$. The truncation is necessary because this does not give complete information about $\alpha$; it only characterizes it up to a certain level.

---
[11] Trunc is truncation, written $\tau$ in Section 3.

In the inm case, we need a code type for x,y and p : P x y and $\alpha$ : inm $p_0$ = inm p. The condition we want is essentially that either $x_0$ = x and $\alpha$ is the determined path inm $p_0$ = inl $x_0$/x = inm p, or $y_0$ = y and $\alpha$ is the determined path inm $p_0$ = inr $y_0$/y = inm p. However, it will be helpful for the remainder of the proof to say this in a bit of a roundabout way. In these two cases, the new information in ((x,y) , p) is either of the form $\Sigma$ (y : Y) P $x_0$ y (when x = $x_0$) or of the form $\Sigma$ (x : X) P x $y_0$ (when y = $y_0$), which suggests taking the disjoint union of these. But $p_0$ itself can have either of these types, and it will be necessary to equate these, so we use the following wedge type, which glues inl ($y_0$ , $p_0$) and inr ($x_0$ , $p_0$) together:

P∨ = Wedge ($\Sigma$ (y : Y) P $x_0$ y) ($\Sigma$ (x : X) P x $y_0$) ($y_0$ , $p_0$) ($x_0$ , $p_0$)

Next, we show that this wedge type does in fact determine a p and a path $\alpha$ : inm $p_0$ = inm p. We define the following map by by pushout elimination:[12]

gluem : P∨ → $\Sigma$ (((x,y) ,p) : Z) (inm $p_0$) = (inm p)
gluem (inl (y , $px_0$y)) = (($x_0$,y) , $px_0$y) , ! (gluel $px_0$y) ○ gluel $p_0$
gluem (inr (x , $pxy_0$)) = ((x,$y_0$) , $pxy_0$) , ! (gluer $pxy_0$) ○ gluer $p_0$
[... coherence given by collapsing inverses ...]

Finally, we define the codes for inm to be the truncated fiber of this map:

CodesFor (inm {x} {y} p) ($\alpha$ : inm $p_0$ = inm p) =
   Trunc i+j (HFiber gluem (((x , y) , p) , $\alpha$)))

For example, we can inhabit this type in the inl case by giving (y, $px_0$y) and then showing that (($x_0$,y) ,$px_0$y) = (x,y,p) (which includes an equation $x_0$ = x, as we said above) and that (along this) $\alpha$ is homotopic to the path ! (gluel $px_0$y) ○ gluel $p_0$ specified in gluem. The advantage of this roundabout presentation is that the type P∨ is exactly the wedge of the two spaces whose connectivity we know by assumption, which will allow us to appeal to wedge connectivity below.

Our definition of CodesFor so far can be summarized as follows. First, we start with functions gluel0 and gluem and gluer0:

```
P x y0              P∨                  P x0 y
  |                  |                    |
  | gluel0 {x}       | gluem              | gluer0 {y}
  ↓                  ↓                    ↓
inm p0=inl x    Σ ( _,p) inm p0=inm p   inm p0=inr y
```

Then, we define each case of CodesFor to be the (i+j)-truncation of the homotopy fiber of the corresponding map, which produces a type that depends on the data in the bottom row (and x, p, y, respectively):

```
CodesFor inl ←--≃--- CodesFor inm ---≃--→ CodesFor inr
    |                      |                    |
    ↓                      ↓                    ↓
inm p0=inl x  ←─────  inm p0=inm p  ─────→ inm p0=inr y
                gluel p              gluer p
```

Now, there are equivalences on the bottom row given by composition with the indicated glue paths, and to complete the definition of CodesFor by pushout-recursion, we now need to construct the two top equivalences, which relate CodesFor inm to CodesFor inl and CodesFor inr up to the corresponding equivalences on the bottom.

---

[12] We use the "thin" pushout for the wedge, so there is no middle constructor, but we still write inl and inr instead of left and right.

*Equivalences* An advantage of the "fat pushout" is that these two cases are symmetric, so we discuss only the case for gluel, which says that the types chosen for inm and for inl agree. More formally, after some standard calculations à la (The Univalent Foundations Program 2013, Ch. 2) and applying univalence, we need to construct an equivalence

Trunc i+j (HFiber gluem (((x , y) , pxy) , $\alpha$m))
$\simeq$ Trunc i+j (HFiber gluel$_0$ $\alpha$l)

given $\alpha$l : Path (inm $p_0$) (inl x) and $\alpha$m : Path (inm $p_0$) (inm pxy) and s : (gluel pxy ○ $\alpha$m) = $\alpha$l, which says that $\alpha$l is in the image of the equivalence indicated in the bottom row of the above diagram. The construction of this equivalence is the most difficult part of the proof.

First, we decompose it into two steps:

Trunc i+j (HFiber (gluel$_0$) $\alpha$l)
$\simeq$ Trunc i+j (HFiber (gluem l pxy) $\alpha$m)        -- Step 1
$\simeq$ Trunc i+j (HFiber gluem (((x,y) , pxy) , $\alpha$))  -- Step 2

where the function gluem l is given a "zig"

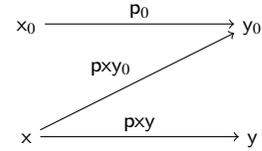

of pxy : P x y and a $pxy_0$ : P x $y_0$ that links it to the fixed $p_0$ : P $x_0$ $y_0$, and produces a path from $p_0$ to pxy in the pushout:

gluem l : ∀ {x y} (pxy : P x y) → P x $y_0$ → (inm $p_0$) = (inm pxy)
gluem l pxy $pxy_0$ = ! (gluel pxy) ○ ! (glue $pxy_0$) ○ gluer $p_0$

This path is the composite

inm $p_0$ = inr $y_0$ = inm $pxy_0$ = inl x = inm pxy

For the first step, first we construct equivalences

HFiber gluel$_0$ $\alpha$l
$\simeq$ HFiber gluel$_0$ (gluel pxy ○ $\alpha$m)
$\simeq$ HFiber (gluem l pxy) $\alpha$m

The first equivalence is given by transporting along s. For the second, because post-composition with a path is an equivalence, we can pull ! (gluel pxy) ○ - into the function part via the hfiber-at-equiv lemma. By expanding the definitions, this gives exactly gluem l. Finally, we can add Trunc i+j to the outside of this equivalence by univalence.

The second step is more complicated. First, both gluem l and gluem are implicitly parametrized by the fixed $p_0$ (we use Agda's section mechanism for this), and gluem l is additionally parametrized by pxy, so by composing with three uses of the aforementioned HFiber-fiberwise-to-total lemma, we can reduce the goal to

Trunc i+j (HFiber gluem l-total ((($x_0$,$y_0$) , $p_0$) , ((x,y) , pxy) , $\alpha$))
$\simeq$ Trunc i+j (HFiber gluem-total ((($x_0$,$y_0$) , $p_0$) , ((x,y) , pxy) , $\alpha$))

where

Z×WZ = $\Sigma$ (((x,y) ,p) : Z) $\Sigma$ (((x',y') ,p') : Z) inm p = inm p'
⟨Z×Z⟩×⟨YX⟩Z = $\Sigma$ (((x,y) ,p) : Z) $\Sigma$ (((x',y') ,p') : Z) P x' y
gluem-total : ($\Sigma$ (( _,p) : Z) P∨ p) → Z×WZ
gluem-total (z , pw) = (z , gluem pw)
gluem l-total : ⟨Z×Z⟩×⟨YX⟩Z → Z×WZ
gluem l-total (z , z' , px'y) = (z , z' , gluem l (snd z) (snd z') px'y)

Recall that Z stands for $\Sigma$ (x,y) P x y and W for Pushout X Y P. We write Z×WZ to suggest the homotopy pullback of two copies of the map ($\lambda$ ( _,p) → inm p) : Z → W, i.e. two copies of Z and a path between them in W. We write ⟨Z×Z⟩×⟨YX⟩Z to suggest the pullback of Z×Z with Z, where the Y component of the first Z is shared with the third, and the X component of the second Z is shared with the third, which reduces to the above type. While gluem and glueml *depend on* $p_0$ (and pxy), these functions act on the total spaces ($\Sigma$-types), accepting *a* $p_0$ and pxy as input, and keeping them unchanged in the output. The reason this change is useful is that it allows us to appeal to the fiber-top-equiv lemma, because we now have two HFibers with the same base. Specifically, this means our goals are

(a) Give a map m-to-ml : ($\Sigma$ ((_,p) : Z) P∨ p) → ⟨Z×Z⟩×⟨YX⟩Z.

(b) Show that it is (i+j)-connected.

(c) Show that it makes the triangle (glueml-total o m-to-ml) = gluem-total commute.

For part (a), the centerpiece is the map from the wedge to the product, which with $x_0$ and $y_0$ and $p_0$ fixed, has type

wedge-to-product : P∨ → ($\Sigma$ (y : Y) P $x_0$ y) × ($\Sigma$ (x : X) P x $y_0$)

and therefore induces a map on total spaces

twtp : $\Sigma$ ((_,$p_0$) : Z) P∨ $p_0$
  → $\Sigma$ ((($x_0$,$y_0$),$p_0$) : Z) ($\Sigma$ (y : Y) P $x_0$ y) × ($\Sigma$ (x : X) P x $y_0$)

The codomain of this map is a rearrangement of ⟨Z×Z⟩×⟨YX⟩Z, just by shuffling components of the tuple:

reassoc-l' :
  $\Sigma$ ((($x_0$,$y_0$) ,$p_0$) : Z) ($\Sigma$ (y : Y) P $x_0$ y) × ($\Sigma$ (x : X) P x $y_0$)
  → ⟨Z×Z⟩×⟨YX⟩Z
reassoc-l' ((($x_0$', y) , px'y), ((y', px'y') , (x , pxy))) =
  (_ , pxy) , (_ , px'y') , px'y

To be able to complete part (c), we also need to precompose with the following on the domain

switchr : ($\Sigma$ ((_,p) : Z) → P∨ p) → ($\Sigma$ ((_,p) : Z) P∨ p)
switchr (z , inl (y,p)) = (z , inl (y,p))
switchr ((($x_0$,$y_0$) ,$p_0$) , inr (x,p)) = ((($x$,$y_0$) ,p) , inr ($x_0$ , $p_0$))
[... coherence given by glue ...]

which leaves the element the same in the inl case, but swaps the inr component with the outer element of Z. Given these, we define

m-to-ml : ($\Sigma$ ((_,p) : Z) P∨ p) → ⟨Z×Z⟩×⟨YX⟩Z
m-to-ml = reassoc-l' o twtp o switchr

Next we need to do part (b), showing that m-to-ml is (i+j)-connected. First, by wedge connectivity, and using the connectivity assumptions about $\Sigma$ (y : Y) P $x_0$ y and $\Sigma$ (x : X) P x $y_0$, the map wedge-to-prod is (i+j)-connected. Next, by the lemma fiberwise-to-total-connected, twtp is (i+j)-connected. Finally, we can show that reassoc-l' and switchr are both equivalences, essentially because they just shuffle data around structurally without losing any. Thus, because the composition of an n-connected map with an equivalence is also n-connected, m-to-ml is (i+j)-connected.

Finally, part (c) is a routine calculation, consisting of a wedge-induction on the domain and some reduction/path algebra. It consists of about 40 lines of code, which would be easy to understand for the interested reader.

This concludes our construction of CodesFor.

### 4.2 Contractibility

Next, we show that the codes fibration is contractible, and that this implies the Blakers–Massey theorem. For most of this section, we assume fixed $x_0$ and $y_0$ and $p_0$ : P $x_0$ $y_0$ as above.

First, we show that CodesFor w $\alpha$ always has an element. To begin, we can construct an element of CodesFor (inm $p_0$) refl by choosing one side of the wedge, and then giving a proof sectioncoh of gluem ($y_0$ , $p_0$) = ((($x_0$,$y_0$) ,$p_0$) , refl), which is just collapsing inverses:[13]

---
[13] We write [ x ] for the proj x constructor of the truncation type.

sectionZ : CodesFor (inm $p_0$) refl
sectionZ = [ inl ($y_0$ , $p_0$) , sectioncoh ]

This extends to an element of CodesFor w $\alpha$ for any w and $\alpha$ : inm $p_0$ = w

section : (w : W) ($\alpha$ : inm $p_0$ = w) → (CodesFor w $\alpha$)
section w $\alpha$ = transport CodesFor' (pair= $\alpha$ connOver) sectionZ

by transporting in CodesFor', which is the uncurrying of CodesFor, with $\alpha$ (and a bit of adjustment for the path position: we have pair= $\alpha$ connOver : (inm $p_0$,refl) = (w,$\alpha$)). This is essentially the same as the path induction we did for this step above, though it is phrased in a way that facilitates the later calculations.

Next, we show that this section is also a retraction, i.e. that when c is a code for $\alpha$, then section $\alpha$ = c. It turns out that it will be enough to do this only for the case where w is inr y:

retraction-r : (y : Y) ($\alpha$ : (inm $p_0$) = (inr y))
  (c : CodesFor (inr y) $\alpha$) → (section (inr y) $\alpha$) = c

This is a rather involved calculation (about 125 lines of code), using truncation induction, path induction, some reductions and path algebra, and the definitions of the equivalences in the cases of CodeFor for gluel and gluer. This step would be simpler in a type theory with better computational properties.

Together, these show CodesFor is contractible for inr:

contr-r : (y : Y) ($\alpha$ : (inm $p_0$) = (inr y))
  → Contractible (CodesFor (inr y) $\alpha$)
contr-r y p = section (inr y) $\alpha$ , retraction-r y $\alpha$

By reducing the pushout elimination defining it, CodesFor (inr y) $\alpha$ is Trunc i+j (HFiber $gluer_0$ $\alpha$). Therefore, it is immediate by expanding the definition of a connected map that

(y : Y) → ConnectedMap i+j ($gluer_0$ {y})

To complete the theorem, we would like to rephrase this as connectivity of the composite glue map that constructs a path inl x = inr y, instead of $gluer_0$. First, by a bit of path algebra, we have

(y : Y) → glue {$x_0$} {y} = ($\lambda$ z → z o ! (gluel $p_0$)) o $gluer_0$

because

   ($gluer_0$ $pxy_0$) o ! (gluel $p_0$)
 = (glue $pxy_0$ o gluel $p_0$) o ! (gluel $p_0$)
 = glue $pxy_0$

Therefore, because composition with a path is an equivalence, and precomposition with an equivalence preserves connectivity, we have

(y : Y) → ConnectedMap i+j (glue {$x_0$} {y})

At the start of this subsection, we assumed $x_0$ and $y_0$ and $p_0$ : P $x_0$ $y_0$. Making these assumptions explicit in the above type, we have now constructed

($x_0$ : X) ($y_0$ : Y) ($p_0$ : P $x_0$ $y_0$) (y : Y)
  → ConnectedMap i+j (glue {$x_0$} {y})

The last step is to discharge the assumption of $y_0$ and $p_0$, to get our overall result

blakers-massey : (x : X) (y : Y) → ConnectedMap i+j (glue {x} {y})

One of our two connectivity assumptions is that for any x, the type $\Sigma$ ($y_0$ : Y) P x $y_0$ is i-connected. By definition, this means that Trunc i ($\Sigma$ ($y_0$ : Y) P x $y_0$) is contractible, so in particular it has an element. We assumed that i is at least −1, and n-connectivity of a map is a (−1)-type (because it is contractibility of something), and therefore it is an i-type, because i ⩾ −1 and the truncation levels can be raised. Therefore, we can eliminate the truncation, and get a $y_0$ and $p_0$ that are related to the given x, producing the inputs to the previous step.

## 5. Conclusion

In this paper, we have described two computer-checked proofs of the Blakers–Massey connectivity theorem using homotopy type theory in Agda. Each mechanization of the theorem itself is about 700 lines of code, though the first is less dense (its character count is about 88% of the second proof), and also uses fewer lemmas from a library. The first proof is more direct in type theory, in the sense that it does not introduce any intermediate higher inductive types, like the wedge type used in the second proof. However, more of the first proof contains more calculations that are difficult to describe in traditional homotopy-theoretic terms. In the second proof, we were better able to describe the construction of the codes equivalences in traditional mathematical language. The proofs that the codes fibrations are contractible are very similar in both cases, though the more mathematical description of the codes equivalences in the second proof results in a more complex calculation at this stage, because there are more intermediate steps to reduce. These two formalizations illustrate the advantages and disadvantages of different ways of working in homotopy type theory.

The result developed in this paper suggests future work both on formalization and on developing new mathematics using the presented techniques. For example, Anel et al. (2016) have generalized the proof of the Blakers–Massey theorem presented here by abstracting the facts about connectedness/truncation involved. Just as the theorem presented here gives isomorphism of homotopy groups in a range, the generalized theorem can be applied to the Goodwillie Calculus to obtain a new result about isomorphism on derivatives in a range. While this concrete application would not be easy to formalize in type theory, we plan to investigate a formalization of the abstract generalized theorem in future work.